\documentclass[twocolumn,showpacs,preprintnumbers,amsmath,amssymb,aps,prl,letterpaper,floatfix,nofootinbib,superscriptaddress,10pt]{revtex4-2}

\usepackage{graphicx}
\usepackage[percent]{overpic}
\usepackage{xcolor}
\usepackage{hyperref}
\hypersetup{colorlinks, linkcolor = [rgb]{0,0.0,0.75}, citecolor = [rgb]{0,0.0,0.75}, urlcolor = [rgb]{0,0.0,0.75}}
\usepackage{bm}
\definecolor{darkgreen}{rgb}{0.0,0.4,0.0}

\newcommand{\I}{\mathrm{i}\mkern1mu}

\begin{document}

\title{\texorpdfstring{$\bm{B\,\to\,\rho\,\ell\,\bar{\nu}}$}{B to rho ell nu} resonance form factors from \texorpdfstring{$\bm{B\,\to\,\pi\pi\,\ell\,\bar{\nu}}$}{B to pi pi ell nu} in lattice QCD }

\author{Luka Leskovec}
\email{luka.leskovec@ijs.si}
\affiliation{Faculty of Mathematics and Physics, University of Ljubljana, Jadranska 19, 1000}
\affiliation{Jo\v zef Stefan Institute, Jamova 39, 1000 Ljubljana, Slovenia}

\author{Stefan Meinel}
\email{smeinel@arizona.edu}
\affiliation{Department of Physics, University of Arizona, Tucson, AZ 85721, USA}

\author{Marcus Petschlies}
\affiliation{Helmholtz-Institut f\"ur Strahlen- und Kernphysik, Rheinische Friedrich-Wilhelms-Universit\"at Bonn, Nußallee 14-16, 53115 Bonn, Germany}

\author{John Negele}
\affiliation{Center for Theoretical Physics, Massachusetts Institute of Technology, Cambridge, MA 02139, USA}

\author{Srijit Paul}
\affiliation{Maryland Center for Theoretical Physics, University of Maryland, College Park, USA}

\author{Andrew Pochinsky}
\affiliation{Center for Theoretical Physics, Massachusetts Institute of Technology, Cambridge, MA 02139, USA}

\begin{abstract}
The decay $B\to\rho \ell\bar{\nu}$ is an attractive process for determining the magnitude of the smallest CKM matrix element, $|V_{ub}|$, and can provide new insights into the origin of the long-standing exclusive-inclusive discrepancy in determinations of this Standard-Model parameter. This requires a nonperturbative QCD calculation of the $B\to\rho$ form factors $V$, $A_0$, $A_1$, and $A_{12}$. The unstable nature of the $\rho$ resonance has prevented precise lattice QCD calculations of these form factors to date. Here, we present the first lattice QCD calculation of the $B\to\rho$ form factors in which the $\rho$ is treated properly as a resonance in $P$-wave $\pi\pi$ scattering. To this end, we use the Lellouch-L\"uscher finite-volume formalism to compute the $B\to\pi\pi$ form factors as a function of both momentum transfer and $\pi\pi$ invariant mass, and then analytically continue to the $\rho$ resonance pole. This calculation is performed with $2+1$ dynamical quark flavors at a pion mass of approximately 320 MeV, and demonstrates a clear path toward results at the physical point.
\end{abstract}
\maketitle

{\it \textbf{Introduction.---}}The Standard Model of particle physics (SM), despite its enormous success, holds unresolved puzzles that point to the existence of new fundamental physics at high energy scales \cite{Fox:2022tzz}. Precision studies of weak decays of hadrons containing bottom ($b$) quarks are a powerful way of searching for the quantum imprints of new interactions beyond the SM \cite{DiCanto:2022icc}. An important part of this program is determining the elements of the Cabibbo-Kobayashi-Maskawa (CKM) quark mixing matrix \cite{Cabibbo:1963yz, Kobayashi:1973fv} through a variety of processes with high precision; inconsistencies between the results from different processes may already point to physics beyond the SM.\\
The smallest and least well-known element of the CKM matrix is $V_{ub}$, which governs transitions from bottom to up quarks. The magnitude $|V_{ub}|$ is determined primarily from semileptonic decays of $b$-hadrons, which have one or more hadrons plus a lepton ($\ell$) and antineutrino ($\bar{\nu}$) in the final state \cite{Gambino:2020jvv}. In the SM, the decay rate of such processes depends on $|V_{ub}|^2$ and on the quantum chromodynamics (QCD) matrix elements of the quark currents $J_V^\mu=\bar{u}\gamma^\mu b$ and $J_A^\mu=\bar{u}\gamma^\mu\gamma_5 b$ between the initial and final hadronic states (the $W$ boson couples only to the ``left-handed'' combination $J^\mu_L = J_V^\mu-J_A^\mu$). By summing the decay rates over all possible up-flavored hadron combinations in the final state, the QCD matrix elements can be obtained through an operator-product expansion and fits to experimental data; this method is known as \emph{inclusive} \cite{Gambino:2020jvv}. Using instead only the rate to a specific hadron or specific combination of hadrons in the final state is known as \emph{exclusive}, and requires nonperturbative QCD calculations of the hadronic matrix elements, which are usually expressed in terms of \emph{form factors}. For a long time, inclusive determinations of $|V_{ub}|$ have yielded significantly higher values than exclusive determinations \cite{Gambino:2020jvv}, and this puzzle remains unsolved. The most precise exclusive determination uses measurements of the $B\to\pi\ell\bar{\nu}$ differential decay rates from \cite{delAmoSanchez:2010af,Lees:2012vv,Ha:2010rf,Sibidanov:2013rkk} and lattice QCD calculations of the $B\to\pi$ form factors \cite{Lattice:2015tia,Flynn:2015mha,Colquhoun:2022atw} and yields $|V_{ub}|^{B\to\pi\ell\bar{\nu}}=3.61(16)\times 10^{-3}$ \cite{FlavourLatticeAveragingGroupFLAG:2024oxs}, while the average of inclusive determinations reported in Ref.~\cite{HFLAV:2022esi} is $|V_{ub}|^{\rm incl.} = 4.19(17) \times 10^{-3}$.

An attractive process for a new exclusive determination of $|V_{ub}|$ is $B\to\rho(770)\ell\bar{\nu}$, where the $\rho(770)$, in the following denoted as just $\rho$, is an unstable resonance with $J^P=1^-$ that decays through the strong interaction to $\pi\pi$. Precise experimental data for this process are available from BaBar \cite{delAmoSanchez:2010af}, Belle \cite{Sibidanov:2013rkk, Belle:2020xgu}, and Belle II \cite{Belle-II:2022fsw,Belle-II:2024xwh}. Furthermore, while the matrix elements of $J_A^\mu$ vanish in $B\to\pi\ell\bar{\nu}$, the process $B\to\rho(\to\pi\pi)\ell\bar{\nu}$ is sensitive to both $J_V^\mu$ and $J_A^\mu$ and provides additional angular observables that can help constrain beyond-the-SM couplings (such as right-handed $J^\mu_R=J_V^\mu+J_A^\mu$ couplings); see Refs.~\cite{Faller:2013dwa,Bernlochner:2014ova,Colangelo:2019axi,Tsai:2021ota,Bernlochner:2021rel,Leljak:2023gna,Bernlochner:2024sfg,Bernlochner:2024ehc} for recent phenomenological studies. The $B\to\rho$ form factors have been calculated at large hadronic recoil (low dilepton invariant mass-squared $q^2$) using light-cone sum rules in the narrow-width approximation \cite{Ball:2004rg,Khodjamirian:2006st,Bharucha:2015bzk,Gubernari:2018wyi}. The full four-body decay $B\to\pi\pi\ell\bar{\nu}$, to which the $\rho$ resonance contributes, was studied using a combination of dispersion relations and heavy-meson chiral perturbation theory in Ref.~\cite{Kang:2013jaa}. The $B\to\pi\pi\ell\bar{\nu}$ form factors have also been calculated using light-cone sum rules at low $q^2$ and low dipion invariant mass, $\sqrt{s}$ \cite{Hambrock:2015aor,Cheng:2017smj,Cheng:2017sfk}, and using QCD factorization at high $\sqrt{s}$ \cite{Boer:2016iez,Feldmann:2018kqr}. A precision determination of $|V_{ub}|$ from $B\to\rho\ell\bar{\nu}$, however, will require a precise lattice QCD calculation of the form factors. The only published lattice calculations of these form factors \cite{UKQCD:1995uhp,Bowler:2004zb,Flynn:2008zr}, which date back more than 20 years ago, were performed in the quenched approximation and treating the $\rho$ as a stable particle. It has since been recognized that a rigorous determination of the $B\to\rho\ell\bar{\nu}$ form factors requires a lattice calculation of the $P$-wave $B\to\pi\pi\ell\bar{\nu}$ form factors as a function of both $q^2$ and $s$, followed by analytic continuation to the $\rho$ resonance pole (and the calculation must be performed with dynamical sea quarks). Following the seminal work by Lellouch and L\"uscher for $K\to\pi\pi$ weak decay \cite{Lellouch:2000pv}, the finite-volume formalism for $1\to 2$ transition matrix elements on the lattice was sufficiently developed to allow its application to $B\to\pi\pi\ell\bar{\nu}$ \cite{Lin:2001ek,Christ:2005gi,Meyer:2011um,Bernard:2012bi,Hansen:2012tf,Briceno:2014uqa,Agadjanov:2014kha,Detmold:2014fpa,Briceno:2015csa}. In the following, we present the first numerical lattice QCD calculation of the $B\to\pi\pi\ell\bar{\nu}$ and $B\to\rho\ell\bar{\nu}$ form factors using this formalism.

{\it \textbf{$\bm{B\to\pi\pi}$ and $\bm{B\to\rho}$ form factors in infinite volume.---}}The transition amplitude of a $B$-meson through the weak flavor-changing current $J^\mu_L$ to $\pi\pi$,
\begin{align}
    &H_{\ell=1}^{\mu}(P,\varepsilon,p_B) \cr &=\langle 
    \pi\pi,\,I=1, I_z=1,\,\ell=1,P,\,\varepsilon \,\,|\,\, J^{\mu}_L \,\,|\,\,B,\,p_B
    \rangle,
    \label{eq:H}
\end{align}
depends on the $B$-meson four-momentum $p_B$, the two-pion four-momentum $P$, and its associated $\ell=1$ partial-wave polarization vector $\varepsilon = \varepsilon(P,m)$, where $m$ is the polarization index. The states are normalized as in Ref.~\cite{Alexandrou:2018jbt}.

We factor the amplitude completely generally as \cite{Briceno:2021xlc}
\begin{align}
    H_{\ell=1}^{\mu}(P,\varepsilon,p_B) = \left(F_V^{\mu}(P,\varepsilon,p_B)-F_A^{\mu}(P,\varepsilon,p_B)\right) \frac{T(s)}{k},
\end{align}
where $T$ is the two-pion scattering amplitude depending on the invariant mass squared $s=P^2$, and $k=\sqrt{s/4-m_\pi^2}$ is the scattering momentum. For an elastic resonance like the $\rho$, the simplest parameterization is a Breit-Wigner function~\cite{Alexandrou:2017mpi}; see the Supplemental Material \cite{Supplemental} for the details. The vector-current component, $F_V^{\mu}$, decomposes using Lorentz symmetry into a single form factor $V(q^2,s)$ as
\begin{align}
F_V^{\mu}(P,\varepsilon,p_B) &= \frac{2 \, \I V(q^2,s)}{m_B + \sqrt{s}} \epsilon^{\mu \nu \alpha \beta} \varepsilon^{\star}_\nu(P,m) P_\alpha (p_{B})_{\beta},
\end{align}
where $q=p_B-P$ and $q^2$ is the lepton-anti-neutrino pair invariant mass squared.
The axial-current component, $F_A^{\mu}$, decomposes into three form factors $A_0$, $A_1$ and $A_{12}$ \cite{Stech:1985tt,Wirbel:1985ji,Horgan:2013hoa,Horgan:2013pva} as
\begin{widetext}
    \begin{align}
&F_A^{\mu}(P,\varepsilon,p_B) =  2 \sqrt{s} A_0(q^2,s) \frac{\varepsilon^\star(P,m) \cdot q}{q^2} q^{\mu}  + A_{12}(q^2,s) \frac{16\; m_B s \; \varepsilon^{\star}(P,m)\cdot q }{\lambda(m_B^2,s,q^2)} \left[ p_B^{\mu} + P^{\mu} - \frac{m_B^2 - s}{q^2}q^{\mu} \right]\cr
&+(m_B + \sqrt{s}) A_1(q^2,s) \left[ \varepsilon^{\star\mu}(P,m) - \frac{\varepsilon^\star(P,m)\cdot q}{q^2}q^\mu -  \varepsilon^\star(P,m)\cdot q \frac{m_B^2 - s -q^2}{\lambda(m_B^2,s,q^2)} \left(p_B^\mu + P^\mu - \frac{m_B^2 -s}{q^2}q^\mu \right) \right].
\end{align}
\end{widetext}
Here, $\lambda$ is the K\"all\'en function. In contrast to the traditional definitions for a $\rho$ final state in the narrow-width approximation, here $m_{\rho}$ is replaced by $\sqrt{s}$ so that in $F_A^{\mu}$ the partially-conserved axial-current relation is manifest. For real values of $s$ and $q^2$, the form factors $f=V,A_0,A_1$ and $A_{12}$ describe the physical process $B\to \pi\pi \ell\bar{\nu}$. To obtain the resonance form factors $f^R$ that describe the contribution of the $\rho$ to the $\pi\pi$ final state, we analytically continue $f$ in the variable $s$ to the complex $\rho$ pole, $\sqrt{s}_{\rho}\approx m_\rho - \tfrac{\I \Gamma_\rho}{2}$, through
\begin{align}
    f^R(q^2)=\frac{c_{\rho}}{k_{\rho}}f(q^2,s=s_{\rho}).
\end{align}
Here, $c_{\rho}^2$ is the residue of the $\rho$ pole in $T$, which near the pole can be written as $T\approx \tfrac{c_\rho^2}{ s_\rho-s}$, and $k_{\rho}$ is the pole location in the complex scattering momentum variable.

\begin{figure*}
    \begin{overpic}[width=0.49\textwidth]{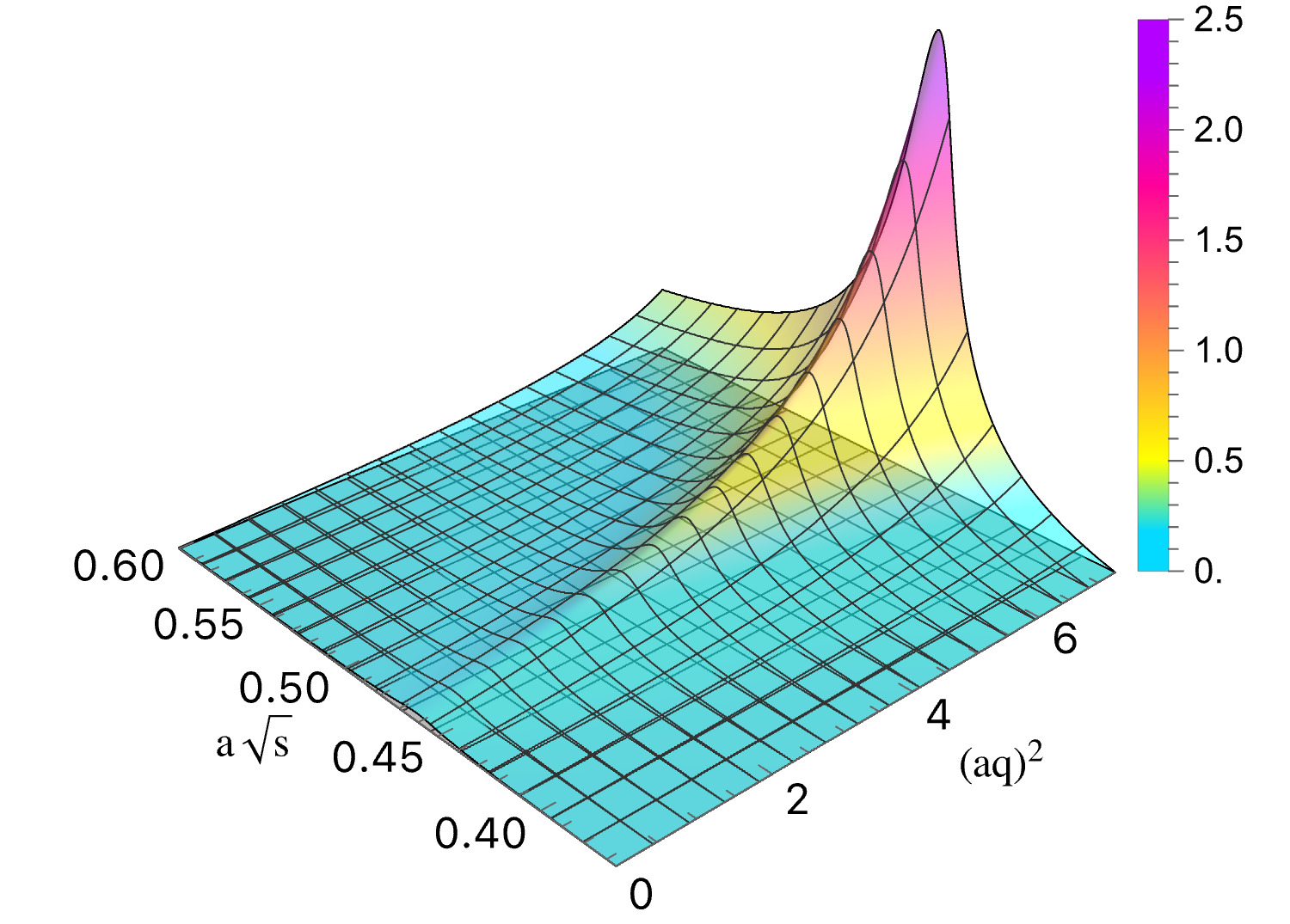}
        \put(5,70){\textbf{(a) $ V(q^2,s) \frac{|T(s)|}{16\pi \sqrt{s}}$}}
    \end{overpic}
    \begin{overpic}[width=0.49\textwidth]{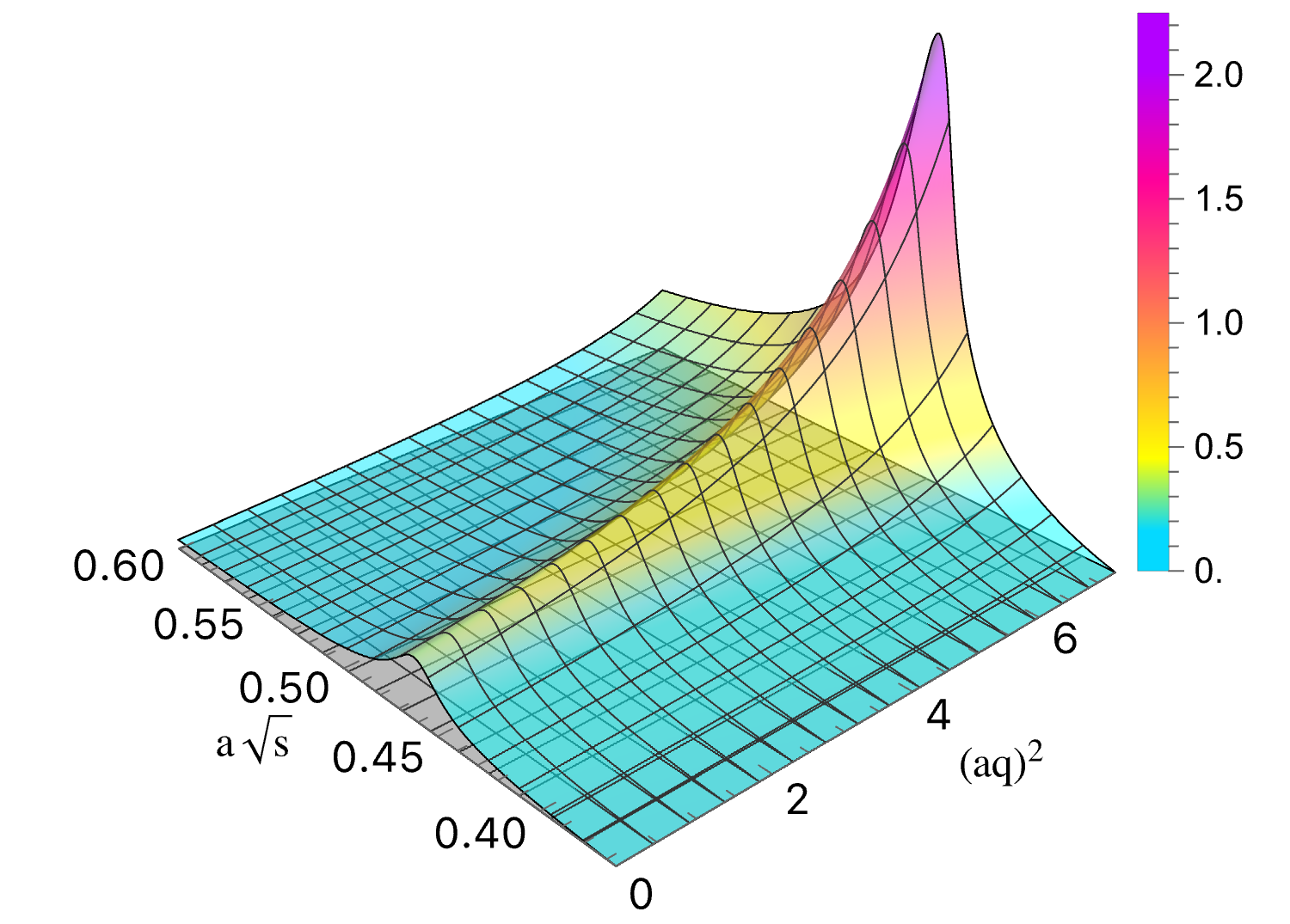}
        \put(5,70){\textbf{(b) $ A_0(q^2,s) \frac{|T(s)|}{16\pi \sqrt{s}}$}}
    \end{overpic} \\
    \begin{overpic}[width=0.49\textwidth]{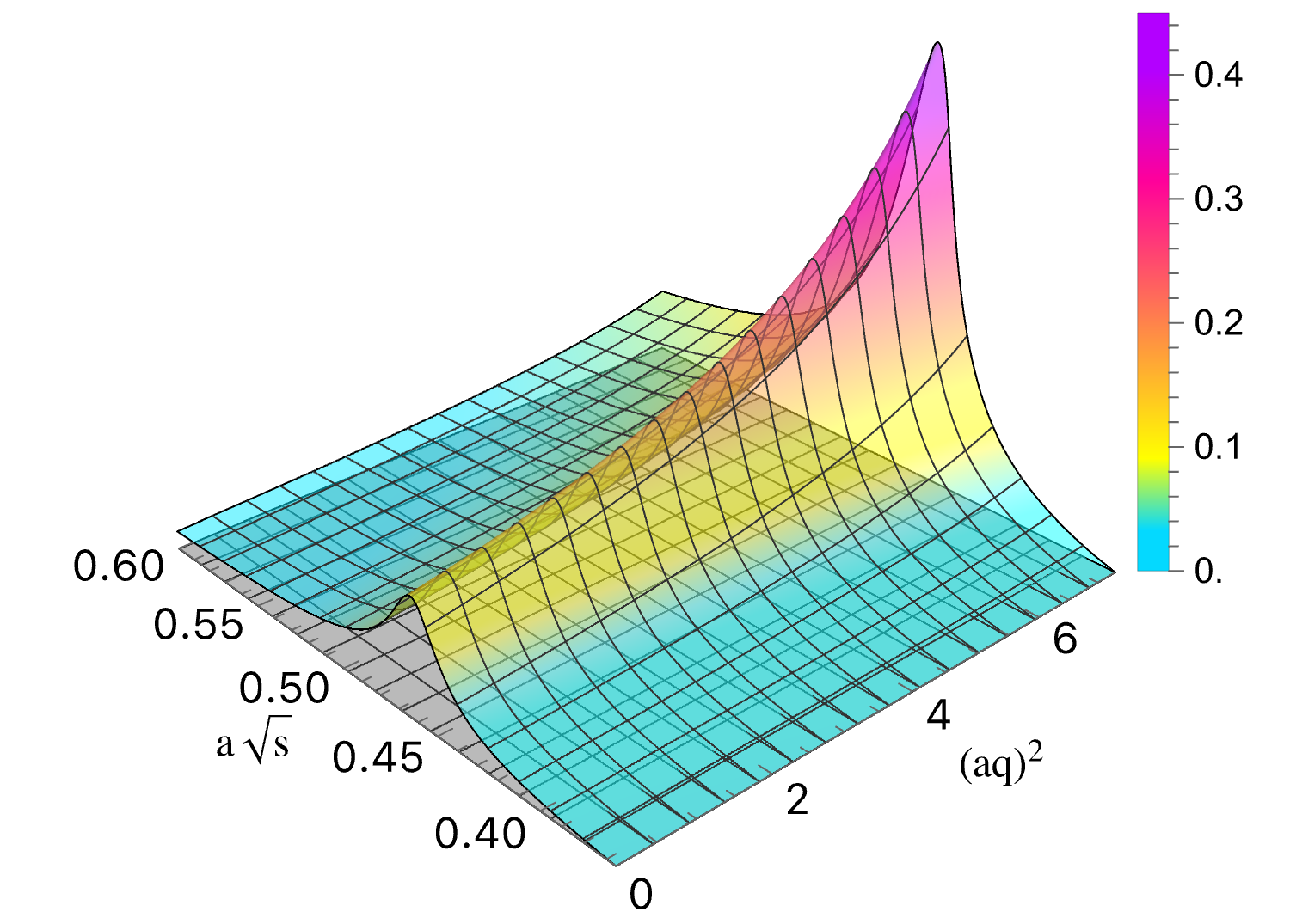}
        \put(5,70){\textbf{(c) $ A_1(q^2,s) \frac{|T(s)|}{16\pi \sqrt{s}}$}}
    \end{overpic}
    \begin{overpic}[width=0.49\textwidth]{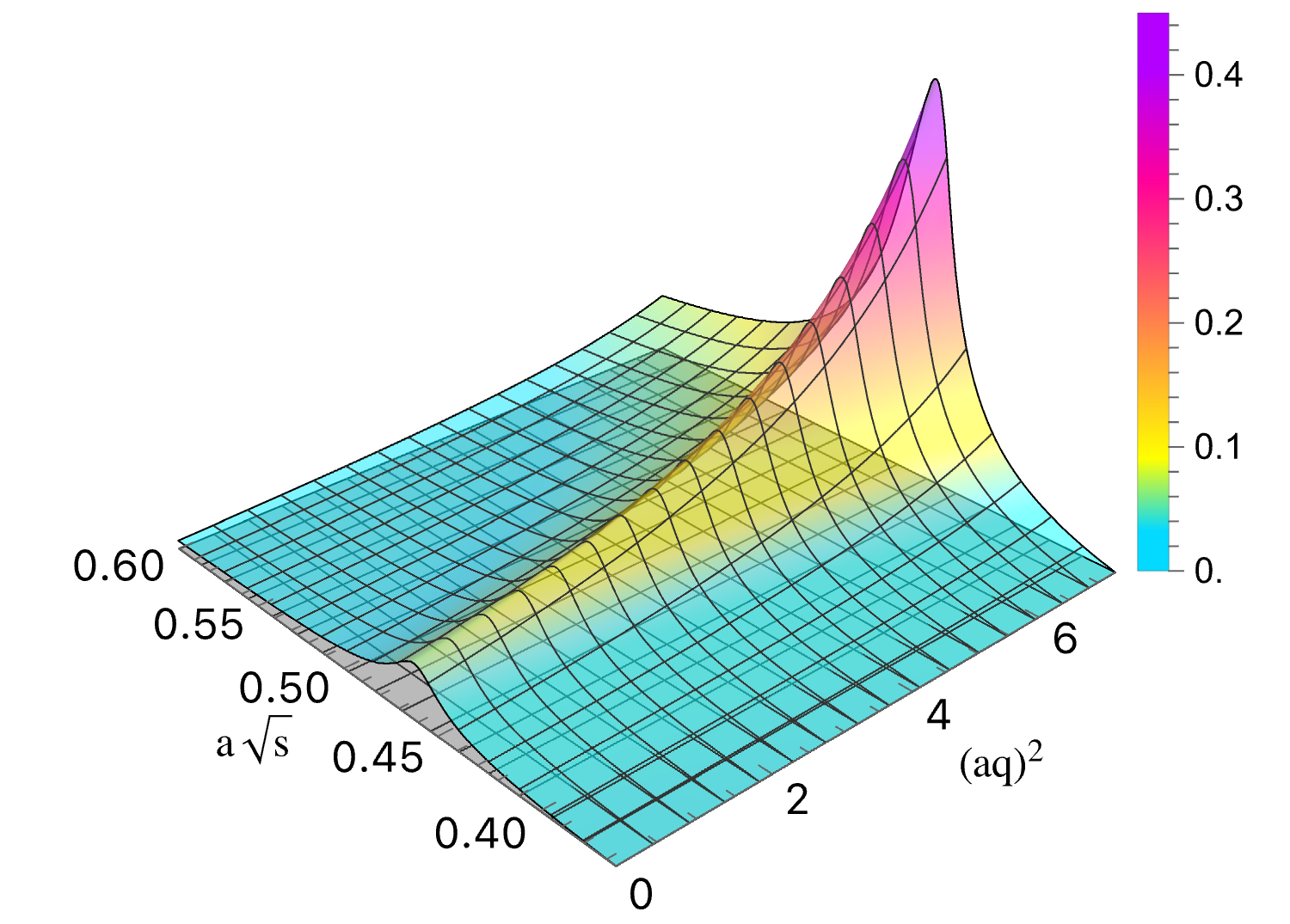}
        \put(5,70){\textbf{(d) $ A_{12}(q^2,s) \frac{|T(s)|}{16\pi \sqrt{s}}$}}
    \end{overpic}
    \caption{Our results for the absolute values of the reduced transition amplitudes $ f(q^2,s) \frac{T(s)}{16\pi \sqrt{s}}$ as a function of $q^2$ and $\sqrt{s}$, for the four different form factors. Here we show the entire $q^2$ region and the $\sqrt{s}$ region in the vicinity of the $\rho$ resonance.} 
    \label{fig:3d}
\end{figure*}

\begin{figure*}
    \includegraphics[width=0.49\textwidth]{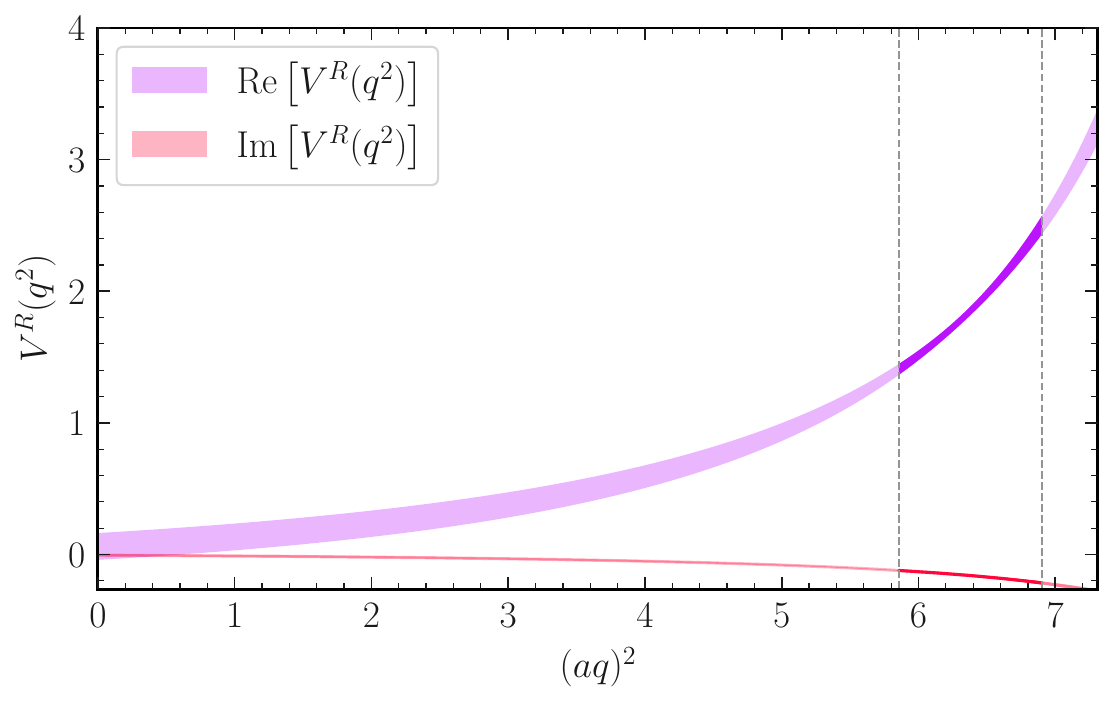}
    \includegraphics[width=0.49\textwidth]{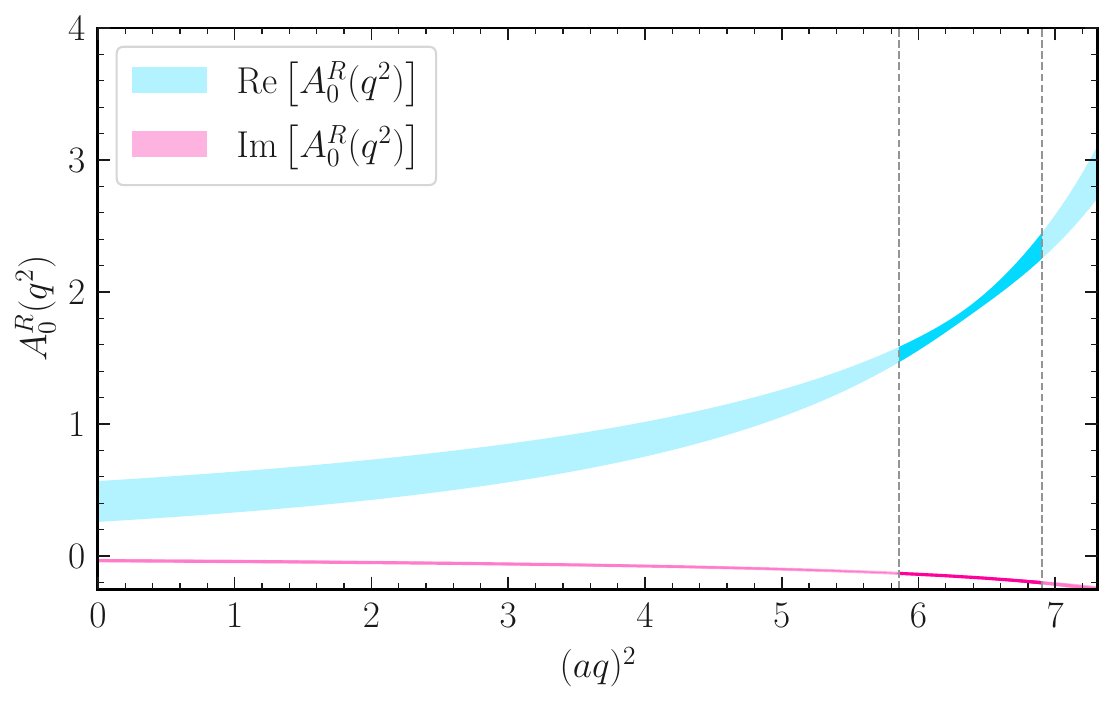}\\
    \vspace{-0.05cm}
    \includegraphics[width=0.49\textwidth]{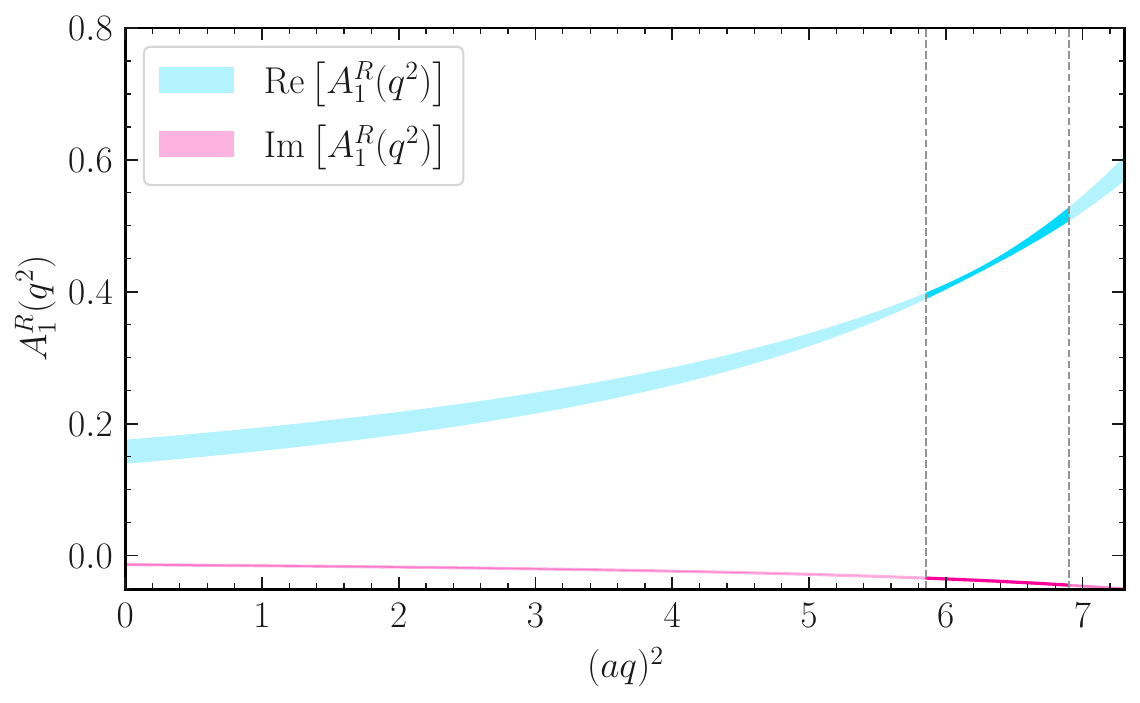}
    \includegraphics[width=0.49\textwidth]{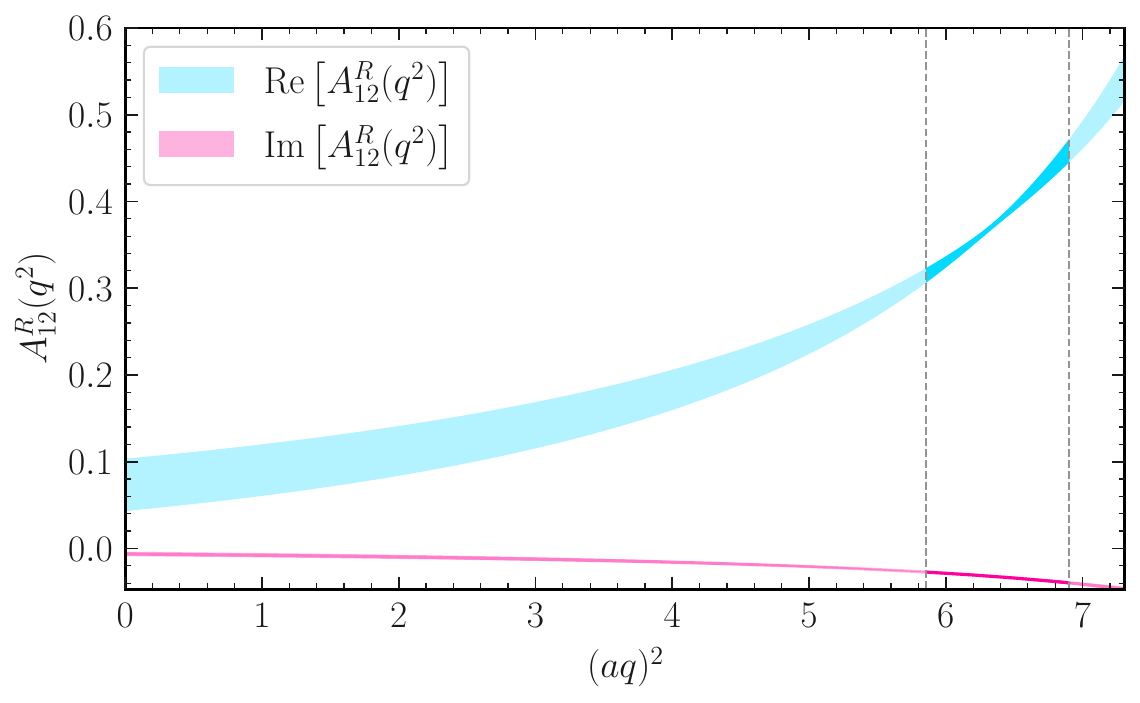}
    \caption{Our results for the resonant form factors $f^R(q^2)$. The dark-shaded region bounded with gray lines is the region in which we have lattice data. The  light-shaded region is an extrapolation based on the first-order $z$ expansion, and the uncertainties at low $q^2$ may be underestimated. \label{fig:resFF}}
\end{figure*}

{\it \textbf{Relation between infinite and finite volume.---}}Lattice QCD calculations evaluate the Feynman path integral in a finite hypercubic box with periodic boundary conditions in space
\begin{align}
    \label{eq:C3}
    &\langle O^{\vec{P},\Lambda,r,n}(\Delta t) J^\mu_L(t_J) O_B^{\vec{p}_B}(0) \rangle \\
    &=\frac{1}{Z} \int \mathcal{D}[U,\psi,\bar{\psi}] O^{\vec{P},\Lambda,r,n}(\Delta t) J^\mu_L(t_J) O_B^{\vec{p}_B}(0)\:e^{-S_{\rm E}}, \hspace{3ex} \nonumber
\end{align}
where $\vec{P}$ labels the two-pion interpolator total momentum at timeslice $\Delta t$, $\Lambda$ the irreducible representation into which the infinite-volume $J^P=1^-$ irrep maps, $r$ the row index of the irreducible representation $\Lambda$, and $n$ the index of the state. The interpolators $O^{\vec{P},\Lambda,r,n}$ are optimized interpolating fields that project to a definite state by using the solutions of the variational problem from the $\pi\pi$ spectroscopy calculation ~\cite{Alexandrou:2017mpi,Alexandrou:2018jbt,Briceno:2015dca,Briceno:2016kkp}. The $B$-meson interpolator at timeslice $0$ always appears in the $A_2^-$ irrep, and we leave the current $J^\mu_L$ at timeslice $t_J$ with the continuum indices. By fitting the time dependence of the three-point function \eqref{eq:C3}, we obtain the finite-volume (FV) matrix elements $\langle \vec{P},\Lambda,r,n|J^\mu_L|p_B\rangle_{FV}$, where $n$ labels the discrete energy eigenstates. These matrix elements are related to the infinite-volume matrix elements through
\begin{align}
    \langle \vec{P},\Lambda,r,n|J^\mu_L|p_B\rangle_{FV} = \sqrt{R^{\vec{P},\Lambda}_n} \langle \vec{P},\Lambda,r,n|J^\mu_L|p_B\rangle, \label{eq:LL1}
\end{align}
where $R^{\vec{P},\Lambda}_n$ is the finite-volume pole residue often referred to as the Lellouch-L\"uscher factor \cite{Lellouch:2000pv,Briceno:2014uqa,Briceno:2015csa}:
\begin{align}
R^{\vec{P},\Lambda}_n= 2E_n   \lim_{E \to E_n} \frac{(E - E_n)}{ \left(F^{\vec{P},\Lambda}\right)^{-1}(s) + T(s)}. \label{eq:LL2}
\end{align}
The Lellouch-L\"uscher factor corrects the normalization and removes finite-volume effects caused by the $\pi\pi$ interactions.
Above, $E_n=\sqrt{s_n+\vec{P}^2}$ is the lattice energy of the $n$-th state in the irrep $\Lambda$, and $F^{\vec{P},\Lambda}$ is the finite-volume function \cite{Briceno:2017max,Luscher:1990ux,Rummukainen:1995vs,Kim:2005gf,Briceno:2014oea}, which is a combination of generalized zeta functions and depends on the spatial lattice size \cite{Leskovec:2012gb}. Equations (\ref{eq:LL1}) and (\ref{eq:LL2}) neglect mixing with higher partial waves, which is very small for the channel considered here \cite{Wilson:2015dqa}.

{\it \textbf{Lattice parameters, analysis, and results.---}}We use a single set of gauge field configurations on a $32^3\times 96$ lattice with a spacing of $a\approx0.11$ fm. This ensemble includes dynamical up, down, and strange Clover-Wilson fermions \cite{Wilson:1974sk, Sheikholeslami:1985ij} with $m_\pi\approx 320$ MeV. We implement the heavy $b$ quark through an anisotropic Clover action \cite{Chen:2000ej,El-Khadra:1996wdx} with parameters tuned to remove heavy-quark discretization errors. Full details of the lattice actions and parameters are given in the Supplemental Material \cite{Supplemental}.

To determine the transition form factors $f=V,A_0,A_1$ and $A_{12}$, on the lattice we calculated three-point correlation functions, \eqref{eq:C3}, using three different source-sink separations and all two-pion irreps with $J^P=1^-$ with total momentum $|\vec{P}| \leq \frac{2\pi}{L}\sqrt{3}$ combined with current momenta insertions with $|\vec{q}|\leq\frac{2\pi}{L}\sqrt{3}$ and $B$-meson momenta with $|\vec{p}_B|\leq\frac{2\pi}{L}\sqrt{3}$. This corresponds to the hiqh-$q^2$ and low-$s$ region, $(aq)^2 \in (5.87,6.91)$ and $a\sqrt{s} \in (0.40, 0.59)$; see the Supplemental Material \cite{Supplemental} for details. The finite-volume matrix elements, $\langle \vec{P},\Lambda,r,n|J^\mu_L|p_B\rangle_{FV}$, are extracted from the three-point functions in two ways: by fitting one- and two-state models for both the source and sink to the raw three-point functions and by fitting one- and two-state models to ratios as in Eq.~(58) of Ref.~\cite{Alexandrou:2018jbt}, which cancels the time dependence up to excited-state contamination. A total of $64$ matrix elements for the vector current, $J_V^\mu=\bar{u}\gamma^\mu b$, and $183$ matrix elements for the axial current, $J_A^\mu=\bar{u}\gamma^\mu\gamma_5 b$, were fitted and model-averaged matrix elements were used in the analysis \cite{Jay:2020jkz}; details are shown in the Supplemental Material \cite{Supplemental}.

We parametrized each of the form factors $f=V,A_0,A_1,A_{12}$ with the $z$-expansion
\begin{align}
    f(q^2,s) = \frac{\sum_n a_n^{(f)} z^n}{1-\tfrac{q^2}{m_P^2}},
\end{align}
where $m_P^2$ is the nearest particle pole in $q^2$ and $z$ is the conformal mapping to the inside of the unit circle \cite{Boyd:1994tt,Boyd:1997qw,Bourrely:2008za,Bhattacharya:2011ah,Bhattacharya:2015mpa,Meyer:2016oeg}:
\begin{align}
\label{eq:zexp}
z = \frac{\sqrt{t_+ - q^2} - \sqrt{t_+ - t_0}}{\sqrt{t_+ - q^2} + \sqrt{t_+ - t_0}}.    
\end{align}
The value of $t_+$ is fixed to the nearest threshold with the correct quantum numbers, while $a^2 t_0=6.0$ for $f=V$ and  $a^2 t_0=6.5$ for $f=A_0,A_1, A_{12}$ is a free parameter chosen such that the data are centered roughly around $z=0$. While, in general, $f$ is a function of both $q^2$ and $s$, we checked various parameterizations and found that $s$-dependence is unnecessary to describe our data.

We follow Ref.~\cite{Briceno:2021xlc} in calculating $R^{\vec{P},\Lambda}_n$, and perform a full jackknife fit using the parameters of the scattering amplitude as determined in our $\pi\pi$ spectroscopy~\cite{Alexandrou:2017mpi}. Our analysis is performed separately for the vector and axial-vector currents. Because only a single form factor appears in the decomposition of the vector matrix elements, the pre-factors can be removed and $V(q^2,s)$ can be fit directly. We  set $a^2 t_+=(am_B+am_\pi)^2$ and $m_P=m_{B^\star}$ because the vector current couples to the $J^P=1^-$ states. Fitting a first-order $z$-expansion yields a $\tfrac{\chi^2}{\rm dof}=\tfrac{72.1}{64-2}=1.16$.

For the axial-current matrix elements, it is not straightforward to isolate each of the three form factors $A_0$, $A_1$, and $A_{12}$.  We instead perform a global fit of all three form factors to the lattice data, using $a^2 t_+=(am_{B^{\star}}+am_\pi)^2$ throughout and $m_P=m_B$ for $A_0$ and $m_P=m_{B_1}$ for $A_1$ and $A_{12}$. Fitting a first order $z$-expansion yields a $\tfrac{\chi^2}{\rm dof}=\tfrac{105.1}{183-5}=0.59$, where we take into account the pole-cancellation condition, $A_0(0)=A_3(0)$ \cite{Stech:1985tt,Wirbel:1985ji}. The results for the $z$-expansion fit and the $\pi\pi$ scattering amplitude along with correlations between all the parameters are given in the Supplemental Material \cite{Supplemental}.

Plots of the transition form factors as a function of $\sqrt{s}$ and $q^2$ are shown in Fig.~\ref{fig:3d}. There, we show the reduced transition amplitudes $\tfrac{k}{16\pi \sqrt{s}} f(q^2,s) \frac{T(s)}{k} = f(q^2,s) \frac{T(s)}{16\pi \sqrt{s}}$ in which we included a phase-space factor of $\tfrac{k}{16\pi \sqrt{s}}$. We observe a $q^2$ dependence as typical for heavy-hadron-to-light-hadron transition form factors and the expected Breit-Wigner behavior in $\sqrt{s}$ associated with the $\pi\pi$ rescattering.

The resonance form factors $V^{R}(q^2)$, $A_0^{R}(q^2)$, $A_1^{R}(q^2)$ and $A_{12}^{R}(q^2)$ obtained through analytic continuation to the $\rho$ pole are shown in Fig.~\ref{fig:resFF}. While the resonance form factors are well-defined, they do not contain the full information on the $B\to\pi\pi \ell\bar{\nu}$ process. We anticipate that future joint experimental and theoretical studies will benefit from utilizing the full dependence on both $\sqrt{s}$ and $q^2$ in the $B\to\pi\pi\ell\bar{\nu}$ amplitude, rather than relying solely on the resonance-pole contribution.

{\it \textbf{Conclusions.---}}Using the rigorous finite-volume formalism for $1\to2$ transition matrix elements, we have performed a lattice QCD calculation of the $B\to \pi\pi \ell\bar{\nu}$ process in which the two-pion final state is produced through the $\rho$ resonance. We have obtained both the two-hadron transition amplitudes and the resonance form factors at a pion mass of $\approx 320 $ MeV. The resonant form factors have a statistical uncertainty of order 2\% in the region where we have data.
Computations at smaller lattice spacings and lighter pion masses are underway, which will enable an extrapolation to the physical point and the application to $|V_{ub}|$. Combined with current and future experimental data, a few-percent total uncertainty on $|V_{ub}|$ from $B\to \rho(\to \pi\pi) \ell\bar{\nu}$ appears within reach.

This work is the first lattice QCD calculation of a semileptonic weak decay with a two-hadron final state. The methodology is widely applicable to several other important processes in particle and nuclear physics, such as $B\to K^\star(\to K\pi)\ell^+\ell^-$.

{\textit{Acknowledgments:}}  We thank Kostas Orginos, Balint Joó, Robert Edwards, and their collaborators for providing the gauge-field configurations. We are grateful to Svjetlana Fajfer and Jonathan Kriewald for valuable discussions. Computations for this work were carried out in part on (1) facilities of the USQCD Collaboration, which are funded by the Office of Science of the U.S.~Department of Energy, (2) facilities of the Leibniz Supercomputing Centre under project pr27yo, which is funded by the Gauss Centre for Supercomputing,  (3) facilities at the National Energy Research Scientific Computing Center, a DOE Office of Science User Facility supported by the Office of Science of the U.S.~Department of Energy under Contract No.~DE-AC02-05CH1123, (4) facilities of the Extreme Science and Engineering Discovery Environment (XSEDE), which was supported by National Science Foundation grant number ACI-1548562, and (5) the Oak Ridge Leadership Computing Facility, which is a DOE Office of Science User Facility supported under Contract DE-AC05-00OR22725. L.L. acknowledges the project was financially supported by the Slovenian Research Agency through projects J1-3034 and N1-0360. S.M.~is supported by the U.S. Department of Energy, Office of Science, Office of High Energy Physics under Award Number D{E-S}{C0}009913. J.N.~and A.P.~acknowledge support by the U.S. Department of Energy, Office of Science, Office of Nuclear Physics under grants DE-SC-0011090 and DE-SC0018121 respectively. A.P.~acknowledges support by the ``Fundamental nuclear physics at the exascale and beyond'' project under grant DE-SC0023116.

\providecommand{\href}[2]{#2}\begingroup\raggedright\endgroup

\end{document}